\documentclass[10pt,letterpaper]{article}
\usepackage{opex3}
\usepackage{graphicx}
\usepackage{cite}
\usepackage{color}
\usepackage{amsmath}
\usepackage{here}
\usepackage{epstopdf}

\newcommand{\be}{\begin{equation}}
\newcommand{\ee}{\end{equation}}
\newcommand{\bea}{\begin{eqnarray}}
\newcommand{\eea}{\end{eqnarray}}

\newcommand\pictc[5]{\begin{figure}
                       \centerline{\vspace{-1mm}
\includegraphics[width=#1\columnwidth,height=0.7\textheight,keepaspectratio]{#3}}
                       \protect\caption{\protect\label{#4} #5}\vspace{-3mm}
                    \end{figure}            }

\newcommand\pict[4][1]{\pictc{#1}{!tb}{#2}{#3}{#4}}
\newcommand\rpict[1]{\ref{#1}}

\begin{document}

\begin{sloppy}

\title{Ultra-directional super-scattering of homogenous spherical particles with radial anisotropy}

\author{Wei Liu$^{\ast}$}
\address{College of Optoelectronic Science and Engineering, National University of Defense
Technology, Changsha, Hunan 410073, China
\email{$^{\ast}$wei.liu.pku@gmail.com}
}

\begin{abstract}
 We study the light scattering of homogenous radially-anisotropic spherical particles.  It is shown that radial anisotropy can be employed to tune effectively the electric resonances, and thus enable flexible overlapping of electric and magnetic dipoles of various numbers, which leads to unidirectional forward super-scattering at different spectral positions.  We further reveal that through adjusting the radial anisotropy parameters, electric and magnetic resonances of higher orders can be also made overlapped, thus further collimating the forward scattering lobes.  The ultra-directional super-scattering we have obtained with individual homogenous radially anisotropic spherical particles may shed new light to the design of compact and efficient nanoantennas, which may find various applications in solar cells, bio-sensing and many other antenna based researches.
\end{abstract}


\ocis{(290.5850) Scattering, particles; (290.4020)  Mie theory; (260.5740) Resonance.}


\section{Introduction}

Recently stimulated by the demonstrations of optically-induced magnetic responses of high primitivity dielectric particles~\cite{Evlykuhin2010_PRB,Garixia_etxarri2011_OE,Kuznetsov2012_SciRep,Evlyukhin2012_NL}, the original proposal by Kerker~\cite{Kerker1983_JOSA} and the concept of Huygens source in the antenna theory~\cite{Love1976_RS,Jin2010_IEEE,Krasnok2011_Jetp}, both of which are related to the manipulation of particle scattering patterns, have attracted renewed surging interest~\cite{Liu2014_CPB, Krasnok2011_Jetp,Liu2012_ACSNANO, Geffrin2012_NC,Filonov2012_APL,Krasnok2012_OE,Rolly2012_OE,Liu2012_PRB081407,Miroshnichenko2012_NL6459,Fu2013_NC,Person2013_NL,Liu2013_OL2621,Staude2013_acsnano,Hancu2013_NL,Vercruysse2013_NL,Liu2014_ultradirectional,Lukyanchuk2014_arXiv_Optimum}. The central precondition for the realization of Kerker's proposal and Huygens source is the efficient excitation of magnetic resonances and their interferences with the electric counterparts~\cite{Kerker1983_JOSA,Love1976_RS,Jin2010_IEEE,Krasnok2011_Jetp,Liu2014_CPB}. Such interferences between electric and magnetic resonances can be employed for efficient shaping of scattering patterns for nanoparticles in the far-field~\cite{Liu2014_CPB,Nieto2010_OE,Gomez-Medina2011_JN,Krasnok2011_Jetp,Liu2012_ACSNANO,Geffrin2012_NC,Filonov2012_APL,Krasnok2012_OE,Rolly2012_OE,Liu2012_PRB081407,Miroshnichenko2012_NL6459,Fu2013_NC,Person2013_NL,Liu2013_OL2621,Staude2013_acsnano,
Hancu2013_NL,Vercruysse2013_NL,Liu2014_ultradirectional,Lukyanchuk2014_arXiv_Optimum}, which thus might paly a critical role in many applications for biological sensing, compact nanoantenna design, photovoltaic devices and so on.

Among all kinds of manipulations of the scattering patterns based on optically-induced magnetic responses,  simultaneous forward scattering enhancement (forward super-scattering) and  backward scattering suppression based on overlapping electric dipoles (EDs) and magnetic dipoles (MDs), which satisfies the first Kerker's condition~\cite{Kerker1983_JOSA,Alu2010_JN}, is one of the most attractive topics~\cite{Liu2014_CPB,Liu2012_ACSNANO,Geffrin2012_NC,Fu2013_NC,Person2013_NL,Liu2013_OL2621}. Though for homogenous dielectric spheres, both EDs and MDs can be excited and actually unidirectional forward scattering with suppressed backward scattering has been experimentally demonstrated~\cite{Geffrin2012_NC,Fu2013_NC,Person2013_NL}. The problem is that the corresponding functioning wavelength is far from the resonant position, due to the fact that the EDs and MDs are spectrally separated, resulting in reduced overall scattering.  To obtain simultaneous unidirectional scattering in the super-scattering regime~\cite{Ruan2010_PRL,Ruan2011_APL,Ali2014_APL_superscattering}, with enhanced forward scattering and sufficiently suppressed backward scattering, basically EDs and MDs have to be spectrally overlapped. This has recently been realised in core-shell plasmonic nanoparticles~\cite{Liu2014_CPB,Paniague2011_NJP,Liu2012_ACSNANO,Liu2013_OL2621}, dielectric nanodisks~\cite{Staude2013_acsnano} or in spheroidal dielectric nanoparticles~\cite{Lukyanchuk2014_arXiv_Optimum}.

Here in this paper, we demonstrate an alternative way based on radial anisotropy to spectrally overlap electric and magnetic resonances, which subsequently leads to unprecedented scattering patterns for individual homogenous particles. It is shown that in spherical dielectric particles, the resonant positions of electric multipoles can be effectively tuned by radial anisotropy, thus enabling flexible overlapping of electric and magnetic resonances of various orders and numbers. Beyond the conventional scattering manipulation to obtain unidirectional forward super-scattering with negligible backward scattering based on overlapped EDs and MDs, we demonstrate that the directionality of the forward scattering can be further improved.  This has been realized by overlapping electric and magnetic resonances of higher orders through optimizing radial anisotropy parameters, which result in ultra-directional forward super-scattering of individual homogenous particles. The mechanism we have revealed of radial-anisotropy induced ultra-directional forward super-scattering can be certainly extended to particles of other kinds of anisotropy~\cite{Stout2006_JOSAA_Mie,Qiu2010_IPOR_Light,Ni2013_OE_controlling} and shapes, and it is expected that such mechanism can shed new light to many particle scattering based applications, especially those which require flexible scattering pattern shaping and control.

\section{Theory of plane wave scattering by radially anisotropic spherical particles}

\pict[0.87]{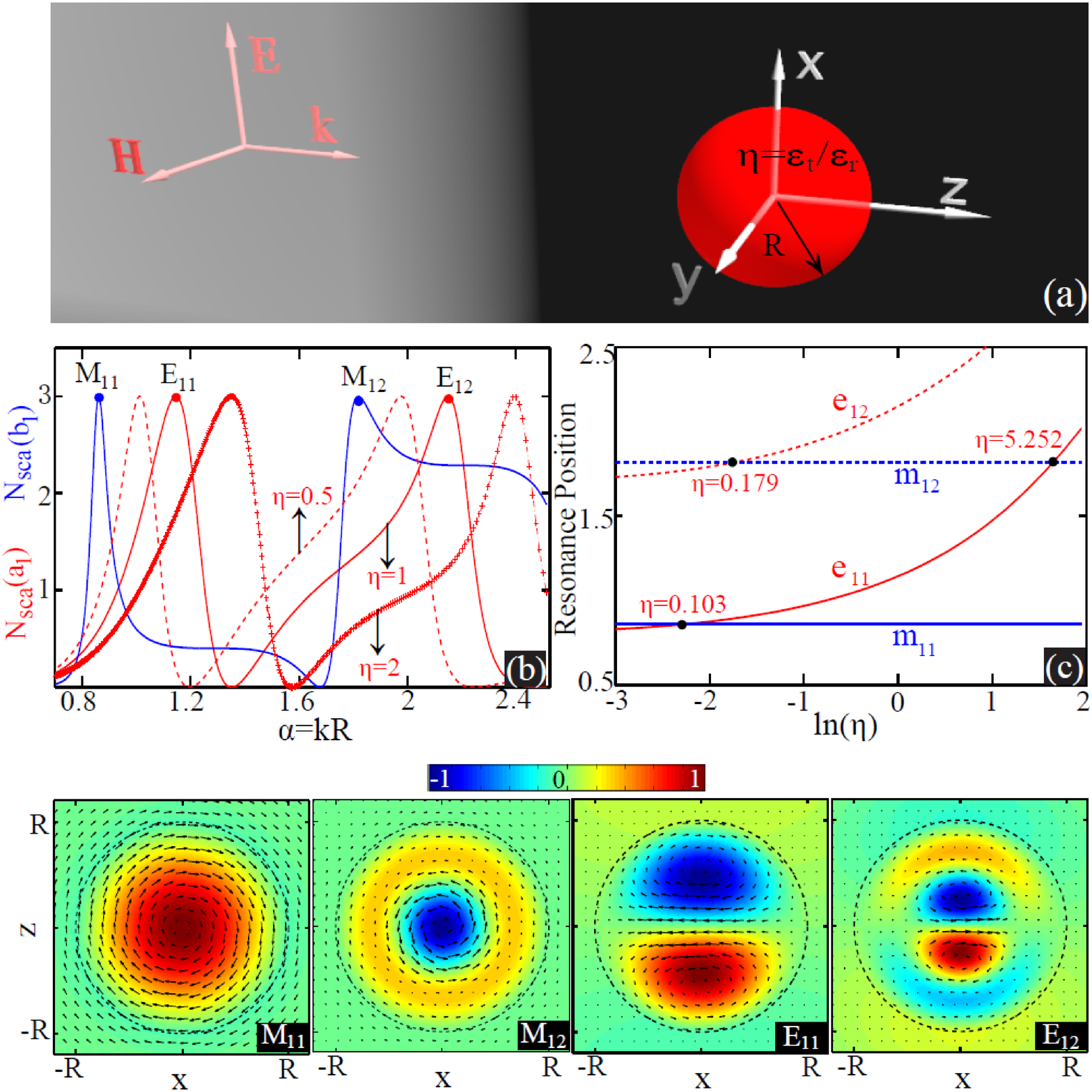}{fig1}{\small (a) Scattering of an incident plane wave by a radially anisotropic spherical particle of radius R, radial permittivity $\varepsilon_r$ and  and transverse permittivity $\varepsilon_t=m^2$. The radial anisotropy parameter is defined as $\eta=\varepsilon_t/\varepsilon_r$. The plane wave is polarized (in terms of electric field) along $x$ direction and is propagating along $z$ direction. (b)  Normalised scattering cross sections for ED [$N_{\rm sca}(a_1)$, red curves] and MD [$N_{\rm sca}(b_1)$, blue curve] of different anisotropy parameters.  The resonance positions of EDs and MDs are indicated by $\textbf{E}_{\rm 11, 12}$ and $\textbf{M}_{\rm 11, 12}$ respectively, with the corresponding near-field distributions at those points [on the $x-z$ plane of $y=0$ in terms of both out of plane $\textbf{H}_y$ (color-plots) and in plane $\textbf{E}$ (vector-plots)] shown in the bottom row. The dashed black curves indicate the boundaries of the particles. (c) The dependence of  resonance positions [solutions of Eq.~(\ref{resonance_positions})] of EDs ($e_{\rm 11, 12}$) and MDs ($m_{\rm 11, 12}$) on the anisotropy parameter $\eta$.}

As is illustrated in  Fig.~\rpict{fig1}(a), we study the scattering of plane waves  by radially anisotropic spherical particles (with radius $R$). The effectively permittivities along radial and transverse directions are $\varepsilon_r$ and  $\varepsilon_t=m^2$ respectively, with the anisotropy parameter defined as $\eta=\varepsilon_t/\varepsilon_r$.  The plane wave is polarized along $x$ direction and propagating along $z$ direction. The scattering can be solved analytically using modified Mie theory~\cite{Bohren1983_book,Qiu2008_JOSAA,Qiu2010_IPOR_Light,Ni2013_OE_controlling} and the total normalised scattering cross section $N_{\rm sca}$ [normalized by ${2\pi \over {{k^2}}}$, where $k$ is the angular wave number in the background (it is vacuum in this study)] is:

\begin{equation}
\label{cross_section}
{N_{\rm sca}} = \sum\limits_{n = 1}^\infty  {(2n + 1)({{\left| {{a_n}} \right|}^2}}  + {\left| {{b_n}} \right|^2}),
\end{equation}
where $a_n$ and $b_n$ are Mie scattering coefficients and for radially anisotropic particles can be expressed as~\cite{Bohren1983_book,Qiu2008_JOSAA,Qiu2010_IPOR_Light,Ni2013_OE_controlling}:

\begin{gather}
a_n  = {{m\psi _{\tilde n} (m\alpha )\psi '_{n} (\alpha ) - \psi _{n} (\alpha )\psi '_{\tilde n} (m\alpha )} \over {m\psi _{\tilde n} (m\alpha )\xi '_{n} (\alpha ) - \xi _{n} (\alpha )\psi '_{\tilde n} (m\alpha )}} \label{an}, \\
b_n  = {{\psi _n (m\alpha )\psi '_n (\alpha ) - m\psi _n (\alpha )\psi '_n (m\alpha )} \over {\psi _n (m\alpha )\xi '_n (\alpha ) - m\xi _n (\alpha )\psi '_n (m\alpha )}} \label{bn},
\end{gather}
where

\begin{equation}
\label{anisotropy}
\tilde n = \sqrt {n(n + 1)\eta  + {1 \over 4}}  - {1 \over 2},
\end{equation}
and $\alpha=kR$; $\psi$ and $\xi$ are Riccati-Bessel functions~\cite{Bohren1983_book} and the accompanying primes indicate their differentiation with respect to the entire argument. Considering that $a_n$ and $b_n$ correspond to \textit{n-th} order electric and magnetic multipoles respectively (\textit{e.g.}, $a_1$ and $b_1$ correspond to ED and MD respectively), then the $N_{\rm sca}$ contributed from them can be expressed respectively as:

\begin{equation}
\label{cross_section_individual}
{N_{\rm sca}(a_n)} =  {(2n + 1){{\left| {{a_n}} \right|}^2}},~~~{N_{\rm sca}(b_n)} =  {(2n + 1){{\left| {{b_n}} \right|}^2}}.
\end{equation}
Since $|a_n|,~|b_n|  \le 1$~\cite{Bohren1983_book,Qiu2010_IPOR_Light}, the maximum $N_{\rm sca}$ of a single \textit{n-th} order multipole (either electric or magnetic) would be $2n + 1$, where  super-scattering is defined as the case when the total $N_{\rm sca}$ is large than this number due to the coexistence to multiple multipoles~\cite{Ruan2011_APL}. The resonance positions of the electric and magnetic multipoles of the \textit{n-th} order are defined as the solutions (in terms of $\alpha$) of the equations below respectively:
\begin{equation}
\label{resonance_positions}
|a_n (e_{\rm nj} )| = 1, ~|b_n (m_{\rm nj} )| = 1,
\end{equation}
where $j$ is the number of the solution [counted from smaller to larger $\alpha$ (from smaller to larger particle radius, or from larger to smaller wavelength)] and thus corresponds to the number of a specific multipole resonance~\cite{Lam1992_JOSAB_Explicit,Johnson1993_JOSAA_Theory,Liu2014_arXiv_Geometric}. For example, $\alpha  = e_{12}$ is the resonance position of the second ED, and we would like to note that most of the previous work only focuses on the case of $j=1$ (resonances supported at largest wavelengths) and neglect electric and magnetic multipoles of higher numbers (supported at smaller wavelengths)~\cite{Evlykuhin2010_PRB,Garixia_etxarri2011_OE,Kuznetsov2012_SciRep,Evlyukhin2012_NL,Liu2014_CPB,Liu2012_ACSNANO,Geffrin2012_NC,Filonov2012_APL,
Krasnok2012_OE,Rolly2012_OE,Liu2012_PRB081407,Miroshnichenko2012_NL6459,Fu2013_NC,Person2013_NL,Liu2013_OL2621,Staude2013_acsnano,
Hancu2013_NL,Vercruysse2013_NL,Liu2014_ultradirectional,Lukyanchuk2014_arXiv_Optimum}.

According to Eq.~(\ref{an})-Eq.~(\ref{bn}), magnetic multipoles (characterized by $b_n$) will be dependent on transverse permittivity  $\varepsilon_t=m^2$ only and will not be affected by the radial anisotropy if $m$ is fixed. In contrast,  electric multipoles (characterized by $a_n$) is dependent on the radial anisotropy [through $\tilde n$; see Eq.~(\ref{anisotropy})] and thus can be tuned.  Those different features of electric and magnetic multipoles can be explained as follows: magnetic resonances correspond to transverse electric-field (TE, electric field is perpendicular to the radial direction) distributions as shown in Figs.~\rpict{fig1}-$\textbf{M}_{\rm 11, 12}$, which means that the change of the radial permittivity would not at all affect the magnetic resonances; Electric resonances however, correspond to transverse magnetic-field (TM, magnetic field is perpendicular to the radial direction and there are both radial and transverse electric field components) distributions, which indicates that the change of the permittivity in the radial direction would influence the electric resonances. Except the explicit expressions of $a_n$ , other expressions for radially anisotropic spherical particles ($b_n$ and those expressed by $a_n$ and $b_n$),   including extinction and absorption scattering cross sections, angular scattering intensities, and the scattering matrix will be the same as those for isotropic spherical particles~\cite{Bohren1983_book,Qiu2010_IPOR_Light}. As a result, all the scattering pattern manipulation approaches, especially those based on interferences of electric and magnetic multipoles~\cite{Liu2014_CPB, Krasnok2011_Jetp,Liu2012_ACSNANO, Geffrin2012_NC,Filonov2012_APL,Krasnok2012_OE,Rolly2012_OE,Liu2012_PRB081407,Miroshnichenko2012_NL6459,Fu2013_NC,Person2013_NL,Liu2013_OL2621,Staude2013_acsnano,Hancu2013_NL,Vercruysse2013_NL,Liu2014_ultradirectional,Lukyanchuk2014_arXiv_Optimum}, can be directly applied for radially anisotropic particles.

\section{Unidirectional super-scattering of radially anisotropic particles with overlapped EDs and MDs}

In Fig.~\rpict{fig1}(b) we show the $N_{\rm sca}$ of both EDs and MDs. Throughout the paper without losing the generality, we fix $m=3.5$ and change the radial permittivity to tune anisotropy parameter $\eta$, and consequently the resonance positions of MDs will be fixed. As described in the section above, there are multiple EDs indicated by $\textbf{E}_{\rm 1j}$ and MDs indicated by $\textbf{M}_{\rm 1j}$ (here we have only shown mode number $j$ up to $2$ only, and EDs and MDs of higher numbers can be supported at larger $\alpha$). The corresponding near-field distributions are shown (on the $x$-$z$ plane with $y=0$) in the bottom row of Fig.~\rpict{fig1}, where the color-plots correspond to the perpendicular magnetic fields along $y$ ($\textbf{H}_y$) and the vector-plots correspond to the electric fields on the $x$-$z$ plane.  It is clear that for MDs there are effectively current (denoted by electric field) circulations with strong magnetic field distribution at the center of the circulation (Figs.~\rpict{fig1}-$\textbf{M}_{\rm 11, 12}$); and for EDs there are effectively linearly polarized currents with negligible magnetic field distribution at the center of the particle (Figs.~\rpict{fig1}-$\textbf{E}_{\rm 11, 12}$). As is also indicated by the near-field distributions, the mode number $j$ basically represents the number of anti-nodes of the fields from the centre to the boundary of the particle, which is highly related to the quantum-mechanical shape resonances~\cite{Lam1992_JOSAB_Explicit,Johnson1993_JOSAA_Theory,Liu2014_arXiv_Geometric}.  We note here that for resonances of the same order $n$, different mode order $j$ can be only recognized from the near-field distributions. In the far-field however, mode of the same order but different mode numbers are indistinguishable.

\pict[0.87]{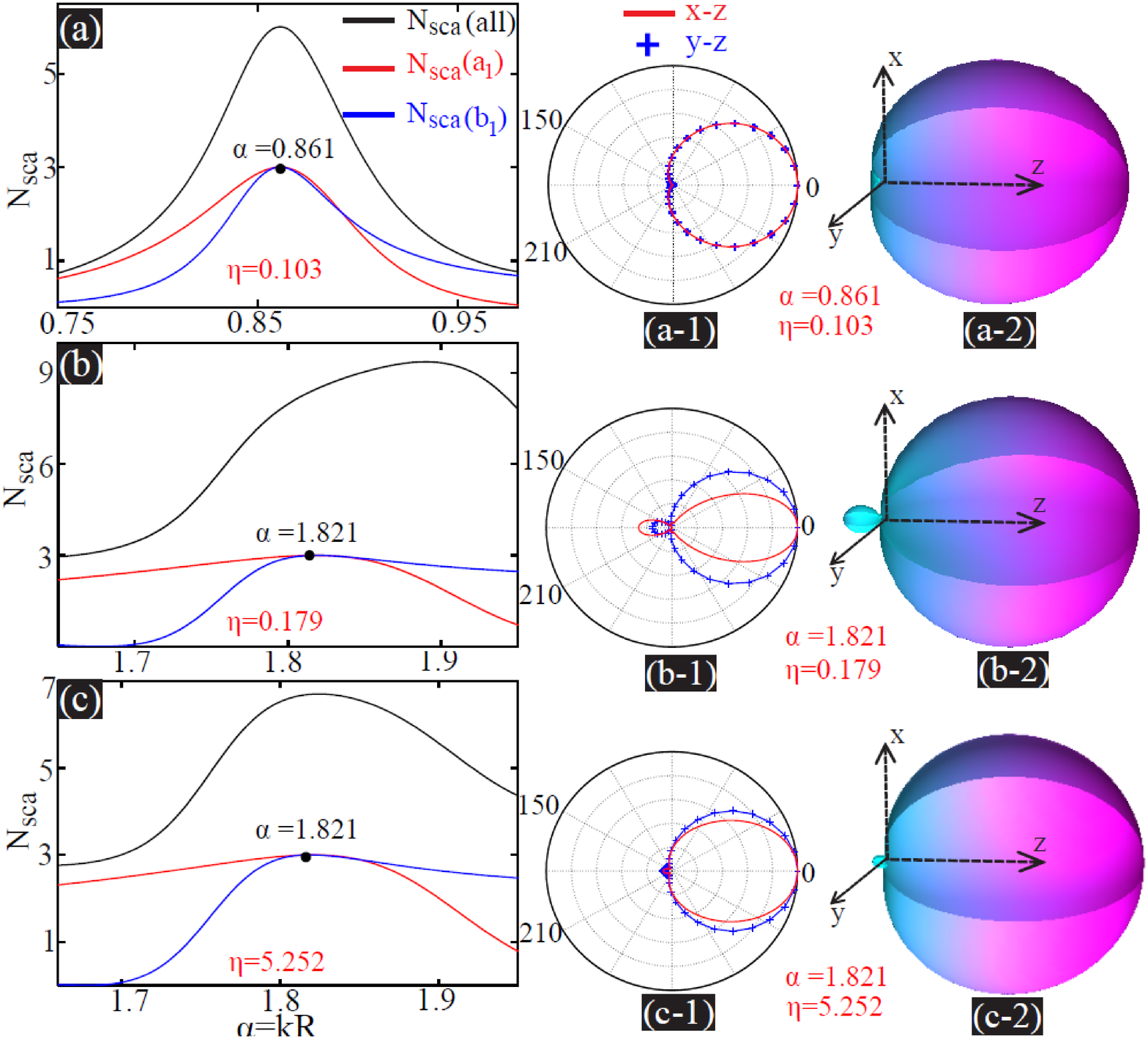}{fig2}{\small Normalised scattering cross section spectra for overlapped EDs (red curves) and MDs (blue curves): (a) $j=1$ for both ED and MD with $\eta=0.103$; (b) $j=2$ for both ED and MD with $\eta=0.179$; (c) $j=1$ for ED and $j=2$ for MD with $\eta=5.252$. For all three cases the total normalised cross section spectra have been provided (black curves) with the overlapping resonant positions of $\alpha=0.861,~1.821,~1.821$ indicated by black dots respectively. The corresponding two-dimensional (red curves: scattering on the $x-z$ planes; blue crosses: scattering on the $y-z$ planes) and three-dimensional (a part is cut off for better visibility) scattering patterns are shown in the middle and right columns respectively.}

Nextly, we study the effect of radial anisotropy on the resonant positions and show the results in Fig.~\rpict{fig1}(b). As has already been described in the section above, with fixed $m$ MD will not be affected by the radial anisotropy, while in contrast ED will red and blue shifted with decreasing and increasing $\eta$ respectively [Fig.~\rpict{fig1}(b)]. This means that basically the relative resonant positions of EDs and MDs can be tuned through change $\eta$, and moreover we can overlap flexibly EDs and MDs even for different mode numbers.

In Fig.~\rpict{fig1}(c) we show the $\eta$ dependence of central resonant positions of EDs ($e_{\rm 1j}$)  and MDs ($m_{\rm 1j}$) for mode number up to $2$ only. Obviously there are three crossing points, where ED and MD are overlapped.  This can be confirmed by their corresponding normalized scattering cross section spectra shown in Figs.~\rpict{fig2}(a)-\rpict{fig2}(c) respectively, where the total cross section has also been shown by black curves.  For all three cases, ED and MD are perfectly resonantly overlapped, and the difference for the three cases is the number $j$ of the modes overlapped.  Previous studies~\cite{Liu2014_CPB,Liu2012_ACSNANO,Geffrin2012_NC,Rolly2012_OE,Liu2012_PRB081407,Fu2013_NC,Person2013_NL,Liu2013_OL2621,Staude2013_acsnano,Liu2014_ultradirectional,Lukyanchuk2014_arXiv_Optimum} focuses only on the case of $j=1$ (resonance at the largest wavelength) while here we have demonstrated that EDs and MDs of higher mode numbers (supported at smaller wavelengths) can be also mode overlapped, providing more freedom for scattering pattern engineering based on interferences between electric and magnetic multipoles.

The corresponding scattering patterns for the three cases at the overlapping resonant positions [indicated by black dots in Figs.~\rpict{fig2}(a)-\rpict{fig2}(c)] are shown in the middle column [two-dimensional scattering patterns on the $x-z$ (red curves) and  $y-z$ (blue crosses) planes] and right column (three-dimensional scattering patters with a part cut off for better visibility) of Fig.~\rpict{fig2}.  For the case of $\eta=0.103$ [Fig.~\rpict{fig2}(a)] at the resonant position ($\alpha=0.861$), since the total $N_{\rm sca}$ (which equals approximately to 6) is about twice of that of the (single-dipole limit, which is $3$), meaning that scattering contributions from other multipoles are negligible. This lead to the more or less ideal case of unidirectional forward super-scattering (with null backward scattering and the scattering pattern is azimuthally symmetric) from overlapped ED and MD~\cite{Kerker1983_JOSA,Liu2014_CPB,Liu2012_ACSNANO}, as demonstrated in Figs.~\rpict{fig2}(a-1) and \rpict{fig2}(a-2). For the other two cases of $\eta=0.179$   [Fig.~\rpict{fig2}(b)] and  $\eta=5.252$  [Fig.~\rpict{fig2}(c)], since the contributions from other multipoles are not negligible (total $N_{\rm sca}$ notably larger than $6$ at the overlapping resonant position of $\alpha=1.821$),  the scattering is not ideally azimuthally symmetric and moreover there are some minor backward scattering [see Figs.~\rpict{fig2}(b-1),~\rpict{fig2}(b-2), \rpict{fig2}(c-1) and \rpict{fig2}(c-2)].   Overall however, for all three cases the backward scattering has been significantly suppressed and forward scattering has been enhanced, justifying our claim of unidirectional forward super-scattering of radially anisotropic particles.

\section{Ultra-directional super-scattering of radially anisotropic particles with overlapped electric and magnetic multipoles of higher orders}

It is recently shown that through overlapping electric and magnetic multipoles of higher orders, ultra-directional forward scattering pattern could be obtained~\cite{Rolly2013_arxiv,Krasnok2013_arxiv,Liu2014_ultradirectional}. Those demonstrations either rely on specific incident dipole sources~\cite{Rolly2013_arxiv,Krasnok2013_arxiv} or  on composite core-shell structures~\cite{Liu2014_ultradirectional}, where the condition of super-scattering cannot always be guaranteed. Here we try an alternative way to employ radial anisotropy to overlap electric and magnetic multipoles of higher orders, thus obtaining ultra-directional forward super-scattering of homogenous particles with plane wave incidence.

\pict[0.87]{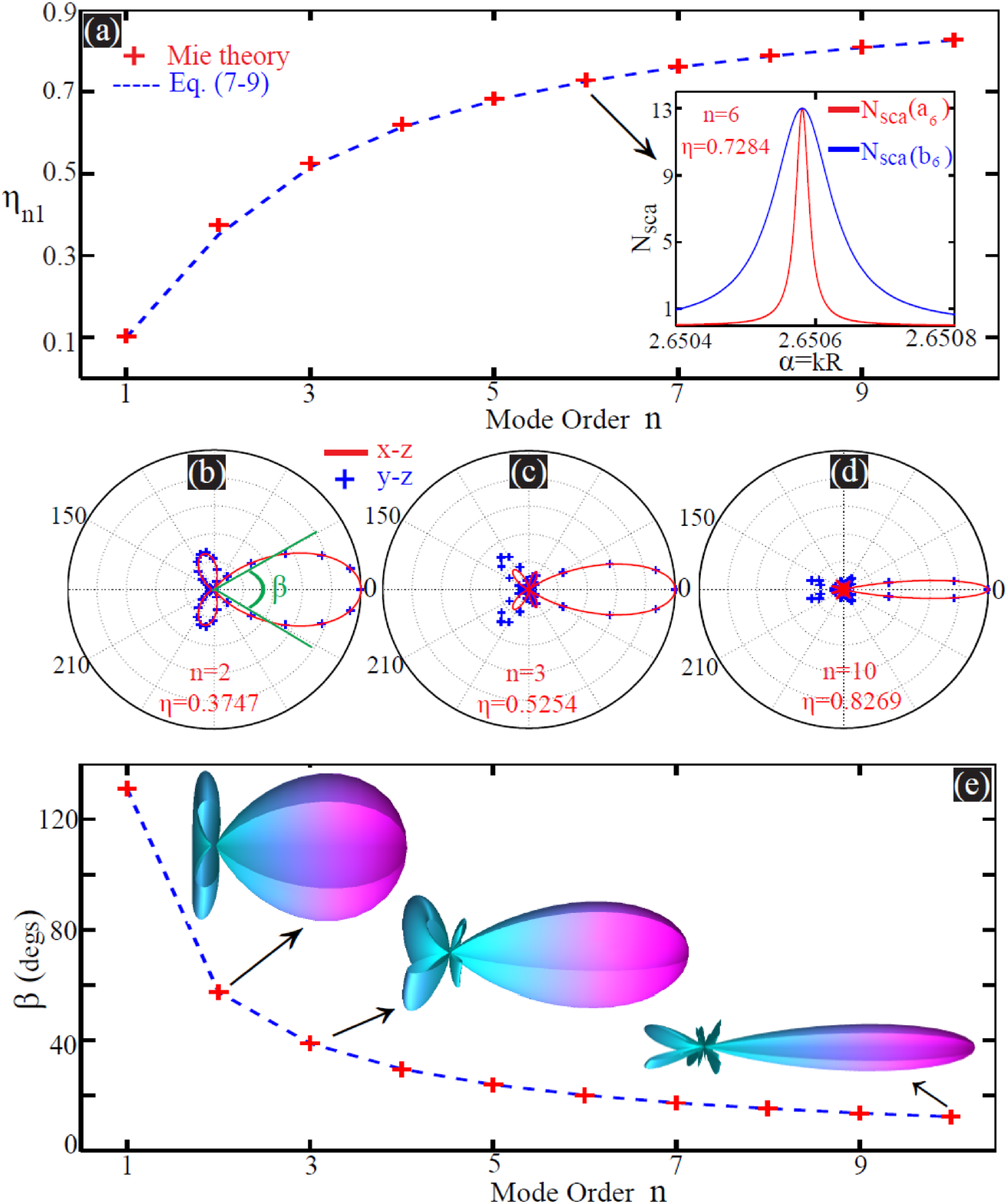}{fig3}{\small (a) The anisotropy parameter required ($\eta_{\rm n1}$) to overlap electric and magnetic multipoles of mode order up to $n=10$ with fixed mode number $j=1$. Both approximated [dashed blue curve, Eq.~(\ref{m_n1})-Eq.~(\ref{overlapping})] and exact results [red crosses, calculated based on Mie theory; see Eq.~(\ref{an})-Eq.~(\ref{resonance_positions})] are provided.  Inset: normalised scattering cross section spectra for overlapped electric (red curve) and magnetic (blue curve) multipoles of $n=6$. Two-dimensional scattering patterns are shown (red curves: scattering on the $x-z$ planes; blue crosses: scattering on the $y-z$ planes) for overlapped electric and magnetic quadrupoles (b), hexapoles (c) and multipoles of $n=10$ (d), where the corresponding anisotropy parameter $\eta=0.3747,~0.5254, 0.8269$ respectively. As is shown in (b), the angular beam-width $\beta$ corresponds to the full width at half maximum of the scattering intensity of the main scattering lobe. (e) The dependence of $\beta$ on the order of the overlapped multipoles and the insets show the three-dimensional scattering patterns of overlapped electric and magnetic quadrupoles, hexapoles and multipoles of $n=10$.}

Without losing generality here we confine the discussion to electric and magnetic multipoles of mode number $j=1$.  First of all to directly evaluate the radial anisotropy parameter $\eta$ required to overlap  electric and magnetic multipoles of different orders, we have extended the simplified expressions for the central resonance positions of EDs and MDs in isotropic particles~\cite{Lam1992_JOSAB_Explicit} to radially anisotropic particles as follows:
\begin{gather}
m_{\rm n1}  = {\chi  \over m} + {{A_1 \chi ^{1/3} } \over {2^{1/3} m}} - {1 \over {(m^2  - 1)^{1/2} }} + {3 \over {10}}{{A_1 ^2 \chi ^{ - 1/3} } \over {2^{2/3} m}} + {1 \over 3}{{m^2 A_1 \chi ^{ - 2/3} } \over {2^{1/3} (m^2  - 1)^{3/2} }}\label{m_n1}, \\
e_{\rm n1} (\eta ) = {{\tilde \chi } \over m} + {{A_1 \tilde \chi ^{1/3} } \over {2^{1/3} m}} - {1 \over {m^2 (m^2  - 1)^{1/2} }} + {3 \over {10}}{{A_1 ^2 \tilde \chi ^{ - 1/3} } \over {2^{2/3} m}} + {1 \over 3}{{A_1 \tilde \chi ^{ - 2/3} } \over {2^{1/3} (m^2  - 1)^{3/2} }}\label{e_n1},
\end{gather}
where $A_1\approx2.338$ is the first root of the Airy function of Ai(-x), $\chi=n+1/2$ and $\tilde \chi=\tilde n+1/2$.  The overlapping of  electric and magnetic multipoles of the $n-th$ order requires that:
\begin{equation}
\label{overlapping}
e_{\rm n1} (\eta _{\rm n1} ) - m_{\rm n1}  = 0,
\end{equation}
where $\eta _{\rm n1}$ is the solution and thus exactly the anisotropy parameter required.

In Fig.~\rpict{fig3}(a) we show $\eta _{\rm n1}$  up to mode order of $n=10$, where both approximated [dashed blue curve, Eq.~(\ref{m_n1})-Eq.~(\ref{overlapping})] and exact results [red crosses, calculated based on Mie theory; see Eq.~(\ref{an})-see Eq.~(\ref{resonance_positions})] are provided. In the inset of Fig.~\rpict{fig3}(a) we show the corresponding $N_{sca}$ spectra for electric and magnetic multipoles of order $n=6$, confirming the ideal multipole overlapping.  The corresponding two-dimensional scattering patters for overlapped quadrupoles ($\eta=0.3747$),  hexapoles ($\eta=0.5254$), and multipoles of $n=10$ ($\eta=0.8269$) are shown in Figs.~\rpict{fig3}(b)-\rpict{fig3}(d) respectively. As has been shown, the backward scattering has been sufficiently suppressed [the minor backward scattering and azimuthal-asymmetry originate from the extra contributions from other multipoles, which is similar to that shown in Fig.~\rpict{fig2}]  and the forward scattering has been more and more collimated with increasing overlapping multipole orders. To characterize quantitatively the forward scattering directionality,  we define the angular beam-width $\beta$ of the main forward scattering lobe , as shown in Fig.~\rpict{fig3}(b), which correspond to the full width at half maximum of the scattering intensity.  We show the dependence of $\beta$ on mode order in  Fig.~\rpict{fig3}(e) and it is clear that as mode order increases, the forward scattering directionality can be further improved. This  confirms the achievement of  ultra-directionality of the scattering from individual homogenous anisotropic particles. This has been further visualized by the insets shown in Fig.~\rpict{fig3}(e), which show three dimensional scattering patterns of overlapped multipoles of $n=2, 3, 10$.

\section{Conclusions and discussions}

To summarise, we study the scattering patterns of radially anisotropic homogenous spherical particles with incident plane waves.  We demonstrate that radial anisotropy can be  employed to tune effectively the central resonance positions of EDs, and thus alow flexible overlapping of EDs and MDs of various mode numbers, which results in unidirectional forward super-scattering patterns with significantly suppressed backward scattering. We further reveal that the forward scattering lobe can be better collimated by optimizing the radial anisotropy parameters to overlap electric and magnetic multipole of higher orders, without compromising the feature of significant backward scattering suppression.

We note that for the overlapping of multipoles of higher orders, we confine our discussions to mode number $j=1$ and such investigations can certainty be extended to multipoles of higher mode numbers, as we have done for dipoles of different mode numbers shown in Fig.~\rpict{fig1} and Fig.~\rpict{fig2}. Also here in this paper we have discussed only radial anisotropy (for natural materials, some anisotropy parameters we have adopted in this paper are not quite practical, which however could probably be realised in artificial metamaterials where a wide range of exotic effective permittivities and permeabilities can be obtained; see \textit{e.g.}, Ref.~\cite{Choi2011_Nature})  and  thus it is natural to expect the employment of anisotropy of other types~\cite{Stout2006_JOSAA_Mie,Qiu2010_IPOR_Light} (including magnetic anisotropy) would render more flexibilities for resonance tuning and thus scattering pattern manipulations. Moreover, when lossy anisotropic materials in included, the effects of loss would be twofold: (1) the resonance positions of both electric and magnetic multipoles would shift~\cite{Liu2014_arXiv_Geometric}, and consequently the anisotropy parameters required to overlap them would be different from those of lossless structures; (2) material loss would change the magnitudes of both electric and magnetic multipoles, which consequently would not necessarily be equal to each other at the overlapping point, and thus the backward scattering cannot be ideally suppressed~\cite{Liu2014_CPB,Liu2012_ACSNANO,Liu2014_ultradirectional}. The scattering shaping based on anisotropic nanoparticles has been discussed in Ref.~\cite{Ni2013_OE_controlling},  which however is contrastingly different what we have presented in this work: (1) They confine their discussions to the quasi-static limit, while our discussions are more general, and can be applied to particles of sizes and functioning spectral regimes that are beyond the quasi-static regime; (2) They confine their discussions to overlapping electric and magnetic dipoles only, while we have discussed overlapping resonances of different orders and numbers, which can further collimate the forward scattering pattern without compromising the feature of backward scattering suppression; (3) Their demonstrations are highly dependent on the magnetic responses and/or magnetic anisotropy,  while our demonstrations are based on electric responses and/or electric anisotropy only and thus relatively more practical.

 Moreover, we should keep in mind that the ultra-directional forward scattering  we have obtained is based on individual particles, and the directionality can be further improved by arranging the anisotropic particles in arrays~\cite{Liu2014_CPB,Liu2012_ACSNANO}. Here we focus on the forward scattering enhancement with backward scattering suppression (namely the first Kerker's condition~\cite{Kerker1983_JOSA,Alu2010_JN}), and radially anisotropy can certainly applied to suppress the forward scattering with enhanced backward scattering (namely the second Kerker's condition~\cite{Kerker1983_JOSA,Alu2010_JN}), or to control the scattering intensities at other scattering angles~\cite{Liu2013_OL2621}. The mechanism we have revealed in this paper is not constrained to spherical homogenous particles and can certainly be extended to particles of other shapes and/or of many layers, which is quite promising for various particle scattering related applications in the fields of nanoantennas, solar cells, bio-sensing and so on.

 \section*{Acknowledgments}

 We thank Andrey E. Miroshnichenko and Jianfa Zhang for valuable discussions.  We also acknowledge the financial support from the National Natural Science Foundation of China (Grant number: $11404403$) and the Basic Research Scheme of College of Optoelectronic Science and Engineering, National University of Defense Technology.

\end{sloppy}
\end{document}